\documentclass[showpacs,amssymb,preprint]{revtex4}
\usepackage{graphicx}
\usepackage{dcolumn}
\usepackage{bm}
\begin{document}
\title{Non-Life Insurance Pricing: Multi Agents Model}
\author{Amir H. Darooneh}
\affiliation{Department of Physics, Zanjan University, P.O.Box
45196-313, Zanjan, Iran.}
\email{darooneh@mail.znu.ac.ir}
\date{\today}
\begin{abstract}
We use the maximum entropy principle for pricing the non-life
insurance and recover the B\"{u}hlmann results for the economic
premium principle. The concept of economic equilibrium is revised
in this respect.
\end{abstract}

\pacs{89.65.Gh, 05.20.-y} \maketitle

\section{Introduction}
Recently the physicists are interested in a new branch of
economics, the insurance market. The financial reaction of
insurance company to variation in number of insurants was studied
by author and his colleague \cite{fd} and subsequently a way for
pricing the insurance premium was suggested by author on the basis
of the equilibrium statistical mechanics \cite{d1,d2,d3,d4}. The
insurer company interacts with it's environment, rest of the
financial market, by money exchanging. In the equilibrium state
the probability for exchanging specified amount of money is
similar to what we see in the canonical ensemble theory.

In this paper we proceed with the viewpoint of the latter
reference for using the principles which are borrowed from
statistical mechanics for premium calculation but restrict
ourselves to a multi agents model for insurance market which was
used previously in the actuary literature \cite{b1,b2}.
\section{B\"{u}hlmann Economic Premium Principle}
The insurance companies and buyers of insurance are the typical
economic agents in the financial market. They compete with each
other to benefit more from their trade. The utility function
demonstrates what an agent is interested for making specified
amounts of profit. The common sense tell us, the agent's utility
function should depend on its financial status which is frequently
described by its wealth, $u(W)$. It is assumed that the utility
function has positive first derivative, $u'(W)>0$, to guarantee
that the agent is willing the profit and negative second
derivative, $u''(W)<0$, to restrict it's avarice. The risk
aversion parameter, $\beta(W)=-u''(W)/u'(W)$, is also involved in
the utility function to scale the agent's will in the market with
respect to it's wealth.

The equilibrium is attained when all agents are satisfied from
their trade. In other word their utility functions should be
maximum in the equilibrium state. This condition may be expressed
as an average form because the risks in the market alter the
agent's wealth randomly.
\begin{equation}\label{e1}
\int_{\Omega} u_{i}(W_{i}(\omega))d\Pi(\omega)=max.
\end{equation}
Where $\omega$ stands for an element of the risk's probability
space $\Omega$. The measure of the integral demonstrates the
weight for occurrence a random event (risk) thus we have,
\begin{equation}
\int_{\Omega} d\Pi(\omega)=1.
\end{equation}
The index $i$ in the eq.\ref{e1} distinguishes the different
agents.

Each agent in the market will be incurred $X_{i}(\omega)$ if
$\omega$ is happening. He insured himself for the price $<Y_{i}>$
and receives $Y_{i}(\omega)$ upon occurrence of this event. The
insurance price is given by,
\begin{equation}\label{ee1}
<Y_{i}>=\int_{\Omega} \varphi(\omega)Y_{i}(\omega)d\Pi(\omega).
\end{equation}
The function $\varphi:\Omega\rightarrow\mathbb{R}$ is called price
density. The agent's wealth also varies due to this trading as
follows,
\begin{equation}
W_{i}(\omega)=W_{0i}-X_{i}(\omega)+Y_{i}(\omega)-<Y_{i}>.
\end{equation}
We suppose the market is a closed system hence the clearing
condition satisfied.
\begin{equation}
\sum_{i}Y_{i}(\omega)=0.
\end{equation}
The sum is over all agents in the market. The above equation in
addition to eq.\ref{e1} let us to find the price density
\cite{b1,b2},
\begin{equation}\label{e2}
\varphi(\omega)=\frac{e^{\beta Z(\omega)}}{\int_{\Omega} e^{\beta
Z(\omega)}d\Pi(\omega)}.
\end{equation}
Where $Z(\omega)$ is aggregate loss in the market,
\begin{equation}
Z(\omega)=\sum_{i}X_{i}(\omega).
\end{equation}
The coefficient $\beta$ comes from combination of risk aversion
parameter of different agents,
\begin{equation}
\frac{1}{\beta}=\sum_{i}\frac{1}{\beta_{i}}.
\end{equation}
The eq.\ref{e2} firstly was derived by B\"{u}hlmann in his famous
articles \cite{b1,b2}. In the following section we retrieve this
result again based on the maximum entropy principle.

\section{The Maximum Entropy Method in Economics}
The risks induce the random condition in the market even the
agents had definite state at the first. The randomness in the
market will be increased when the time goes forward. Finally the
market falls into a state with the most randomness. This is what
we nominate as the equilibrium state. The consequence of
randomness in the market is losing the information about the
agents and their strategies for trading.

The main question in the economics is how we can calculate the
probability for the market to have a specified amount of money. In
other word we look for the probability of happening an event,
namely $\omega$, when the market is in equilibrium. As is seen in
the eq.\ref{ee1} the insurance price is defined in respect to this
probability function, $\varphi(\omega)$, which is called the price
density in the actuary terminology.

The maximum entropy principle appears as the best way when we make
inference about an unknown distribution based only on the
incomplete information \cite{j}. We adopt this method for
calculation the mentioned probability density.

The entropy functional can be written as \cite{j},
\begin{equation}\label{e3}
H[\varphi]=-\int_{\Omega}
\varphi(\omega)\ln\varphi(\omega)d\Pi(\omega).
\end{equation}
The price density should satisfies in normalization condition.
\begin{equation}\label{e4}
\int_{\Omega} \varphi(\omega)d\Pi(\omega)=1.
\end{equation}
The wealth of market is defined as the sum of the agents wealth.
\begin{equation}\label{ee4}
W(\omega)=\sum_{i}W_{i}(\omega).
\end{equation}
We assume that the average of market's wealth is constant.
\begin{equation}\label{e5}
<W>=\int_{\Omega} \varphi(\omega)W(\omega)d\Pi(\omega)=Const.
\end{equation}
In the equilibrium the entropy eq.\ref{e3} has maximum value and
the constrains eqs.\ref{e4} and \ref{e5} should also be satisfied
simultaneously. This mathematical problem can be solved
immediately by the method of Lagrange's multipliers.
\begin{equation}
\delta H[\varphi]+\lambda \delta\int_{\Omega}
\varphi(\omega)d\Pi(\omega)+\beta \delta<W>=0.
\end{equation}

The canonical distribution is the solution to above equation as we
have seen before for special cases that are given in every
statistical mechanics textbooks \cite{p}.
\begin{equation}\label{e6}
\varphi(\omega)=\frac{e^{-\beta W(\omega)}}{\int_{\Omega}
e^{-\beta W(\omega)}d\Pi(\omega)}.
\end{equation}

The above result is derived previously for agent-environment model
of financial market \cite{d4}.

The eq.\ref{e6} may be applied to all economic systems. The
parameter $\beta$ is positive to ensure that extreme values for
market's wealth have small probability.

In the case of insurance the market's wealth is given as'
\begin{equation}
W(\omega)=W_0-Z(\omega)=\sum_iW_{0i}-\sum_iX_i(\omega).
\end{equation}
Since the market's initial wealth is constant then we obtain the
same form for the price density as what is seen in the
eq.\ref{e2}. It is worth to mention that the total risk must be
lesser than the market's initial wealth.

The premium that the $i-$th agent pays is calculated by using the
eq.\ref{e2}.
\begin{equation}
p_i=\frac{\int_\Omega X_i(\omega)e^{\beta
Z(\omega)}d\Pi(\omega)}{\int_{\Omega} e^{\beta
Z(\omega)}d\Pi(\omega)}.
\end{equation}

If the risk function for different agents have no correlation and
dependency then the Esscher principle is obtained \cite{b1,b2}.

\begin{eqnarray}
p_i&=&\frac{\int_{\Omega} X_i(\omega)e^{\beta
X_i(\omega)}d\Pi_i(\omega)\int_{\Omega}e^{\beta
(Z_(\omega)-X_i(\omega))}d\Pi_{other}(\omega)}{\int_{\Omega}
e^{\beta X_i(\omega)}d\Pi_i(\omega)\int_{\Omega}e^{\beta
(Z_(\omega)-X_i(\omega))}d\Pi_{other}(\omega)}\nonumber \\
 &=&\frac{\int_{\Omega} X_i(\omega)e^{\beta
X_i(\omega)}d\Pi_i(\omega)}{\int_{\Omega} e^{\beta
X_i(\omega)}d\Pi_i(\omega)}.
\end{eqnarray}
Where the $d\Pi_i(\omega)$ and $d\Pi_{other}(\omega)$ demonstrate
the weight for the $i-th$ agent risks and remaining part of the market.
The parameter $\beta$ has important roll in price density,
it can be calculated on basis of the method that is introduced in previous works
\cite{d2,d4}, but our intuition from similar case in thermal physics tell us \cite{p},
\begin{equation}
\beta\approx\frac{1}{<W>}.
\end{equation}
This result shows that the wealthier market offers low price and
the prices also depends on the size of risks in the market.

The adopted way for premium calculation is more general and
independent of the market's  models. It also enable us to apply
easily any other constraint which are in the market.

The method of maximum entropy may be generalized for local finite
market via Tsallis definition of entropy. This case is under
investigation.
\begin{acknowledgments}
The author acknowledge Dr. Saeed for reading the manuscript and
his valuable comments. This work has been supported by the Zanjan
university research program on Non-Life Insurance Pricing No.
8234.
\end{acknowledgments}

\bibliographystyle{}

\end{document}